\documentstyle[11pt,epsfig]{article}
\begin{document}
\title{\bf FLRW Cosmology of Induced Dark Energy Model and Open Universe}
\author{Amir F. Bahrehbakhsh\thanks{email: amirfarshad@gmail.com}\\
 {\small Department of Physics, Faculty of Science, Payam-e-Noor University, Iran}}
%
%
\maketitle
\begin{abstract}

We investigate the FLRW type cosmology of the Induced Dark Energy model and
illustrate that the extra terms emerging from the fifth dimension can play the role of dark energy.
The model predicts the expansion with deceleration at early time and acceleration in late time for
an open universe.

\end{abstract}

{\small Keywords: Induced--Matter Theory; FLRW Cosmology; Dark Energy; Geometry of the Universe.}
\bigskip
\section{Introduction}
\indent

Observations of distant type Ia--supernovas
indicate that universe is currently accelerating in its expansion
~\cite{Perlmutter}--~\cite{Bernardis2002}.
Therefore, in a school of thought, the universe should mainly be filled with
what usually is called dark energy.
To explain this acceleration,
a considerable amount of work has been performed in the
literature~\cite{Carroll2001}--~\cite{Yoo2012}.

The first and simplest explanation for dark energy is vacuum energy which,
was introduced for the first time, by Einstein as cosmological constant to explain static universe.
However, after Hubble demonstration of the universe expansion,
Einstein called the cosmological constant his biggest blunder,
but this term is preserved today to explain the current acceleration of the universe expansion~\cite{Carroll2001}.
The cosmological constant has perfect fit with observations and has equation of state
parameter, $w_{_{\Lambda}}=-1$. Though, it suffers two cosmological problems; The theoretical
expectation of dark energy density value is 120 orders of magnitude bigger than the observations
and more over, dark energy density and all matter (the baryonic and dark) energy
density are of the same order today.
Hence, many other models involving non--constant dark energy have been proposed, such
as the quintessence~\cite{QuintessenceA}--~\cite{QuintessenceD}, the k-essence~\cite{K-essenceA,K-essenceB}
and the chaplygin gas~\cite{ChaplyginA,ChaplyginB} models.
However, in most of them presuppose minimally coupled scalar fields with different \emph{priori}
added by hand potentials which, their origins are~not clearly known.

Recently, some efforts based on the Brans--Dicke theory in which, the
scalar field is non--minimally coupled to the curvature, has been performed to explain
today's accelerating expansion of the universe~\cite{Aguilar}--~\cite{Amirfarshad2013}.

On the other hand, most of the attempts for geometrical unification of gravity with
other interactions are based on using higher dimensions addition to
our conventional four--dimensional ($4D$) space--time.
After Nordstr{\o}m~\cite{Nordstrom}, who was the first established a
theory based on extra dimensions, Kaluza~\cite{Kaluza} and Klein~\cite{Klein}
built a five--dimensional ($5D$) version of general relativity (GR) in which
electrodynamics rises from an extra fifth dimension.
After that, an intensive amount of works have been focused on this regard
either via different mechanism for compactification of extra
dimension or generalizing it to non--compact scenarios~\cite{OverduinWesson1997}
such as the space--time--matter (STM) or induced--matter
(IM) theory~\cite{WessonA,WessonB} and the Brane World scenarios~\cite{Pavsic}.
The significant of the IM theory is that inducing $5D$ field equations without
matter sources leads to the $4D$ field equations with matter sources.

In this work, following the Induced Dark Energy (IDE) model introduced in Ref.~\cite{Amirfarshad2016},
for a $5D$ version of GR, with generalized $5D$ Friedmann--Lema\^{\i}tre--Robertson--Walker (FLRW) metric,
we investigate the cosmological implications of the model for an open universe.
For this purpose, in the next section, with a brief review of the IDE model,
we rewrite the cosmological equations and focus on the case of an open universe.
We also provide a table and a figure for a better view of the deceleration and acceleration epoches
of the universe expansion, which specify by the model.
Finally, in the Sec.~$3$ conclusion is presented.

\section{FLRW cosmology of Induced Dark Energy model}
\indent
According to the Ref.~\cite{Amirfarshad2016} we consider $5D$ version of the GR action as
\begin{equation}\label{5D Action}
\emph{S}
=\int\sqrt{|{}^{_{(5)}}g|} \left (\frac{1}{16\pi G}\
^{^{(5)}}\!R+
L_{m} \right )d^{5}x\, ,
\end{equation}
where, $c=1$,
$^{^{(5)}}R$ is $5D$ Ricci scalar, $^{_{(5)}}g$ is the determinant
of $5D$ metric $g_{_{AB}}$ and
$\emph{L}_{m}$ represents the matter Lagrangian.
Therefore, the Einstein field equations in five dimension will be
\begin{equation}\label{5D Einstein Eq}
^{^{(5)}}G_{_{AB}}=8\pi G \
^{^{(5)}}T_{_{AB}} \, ,
\end{equation}
in which, the capital Latin indices run from zero to four, $^{^{(5)}}G_{_{AB}}$ is $5D$ Einstein tensor and
$^{^{(5)}}T_{_{AB}}$ is $5D$ energy--momentum tensor.
As an assumption, we consider $^{^{(5)}}T_{\alpha\beta}$ components of the  $^{^{(5)}}T_{_{AB}}$ to be
the baryonic and dark matter source in a $4D$ hypersurface, i.e. $T^{^{(M)}}_{\alpha\beta}$,
where the Greek indices go from zero to three.
Hence, in this case we take $^{^{(5)}}T_{_{AB}}= diag (\rho_{_{M}},-\emph{p}_{_{M}},-\emph{p}_{_{M}},-\emph{p}_{_{M}},0)$,
in which $\rho_{_{M}}$ and $p_{_{M}}$ are respectively the energy density and the pressure of the matter (baryonic and dark matter).
For cosmological purposes we restrict the $5D$ metrics to the warped ones of the form, in local coordinates $x^{A}=(x^{\mu},y)$,
\begin{equation}\label{5D Metric}
dS^2=\
^{^{(5)}}\!g_{\mu\nu}(x^{C})dx^{\mu}dx^{\nu}+g_{_{44}}(x^{C})dy^2
\equiv \ g_{\mu\nu}dx^{\mu}dx^{\nu}+\epsilon
b^{2}(x^{C})dy^2\, ,
\end{equation}
where $y$ represents the fifth coordinate and $\epsilon^2=1$.
Hereupon, after some manipulations, Eq. (\ref{5D Einstein Eq}) on the hypersurface $\Sigma_{\circ}$ can be
written as
\begin{equation}\label{4D Einstein Eq}
G_{\alpha\beta}=8\pi G(T^{^{(M)}}_{\alpha\beta}+T^{^{(X)}}_{\alpha\beta}),
\end{equation}
where $T^{^{(X)}}_{\alpha\beta}$ is considered as dark energy
component of the energy--momentum tensor~\cite{Amirfarshad2016}.

For a $5D$ universe with an extra
space--like dimension in addition to the three usual
spatially homogenous and isotropic ones, metric (\ref{5D Metric}), as a generalized FLRW solution,
can be written as~\cite{Amirfarshad2016}
\begin{equation}\label{FLRW Metric2}
dS^2=-dt^2+a^2(t)l^2(y)\left [\frac{dr^2}{1-kr^2}+r^2(d\theta^2+\sin^2\theta
d\varphi^2)\right ]+b^2(t)dy^2\, .
\end{equation}
In general, we consider the scale factors to be functions of cosmic time and extra dimension coordinate.
But, for mathematical simplicity, we assume that they are separable functions of cosmic time
and extra dimension coordinate. Besides, the functionality of the scale
factor of the fifth dimension on the extra dimension coordinate can be eliminated by transforming to a
new extra coordinate.

By assuming $H\equiv\dot{a}/a$, $B\equiv\dot{b}/b$ and $L\equiv{l'}/l$ and
considering metric (\ref{FLRW Metric2}), the Einstein Eqs. (\ref{5D Einstein Eq}) reduce as follows~\cite{Amirfarshad2016};
\begin{equation}\label{FLRW1}
H^2=\frac{8\pi G}{3}\tilde{\rho}-\frac{k}{a^2 l^2}\, ,
\end{equation}
\begin{equation}\label{FLRW2}
\frac{\ddot{a}}{a}=-\frac{4\pi G}{3} (\tilde{\rho}+3\tilde{p})\,
\end{equation}
and
\begin{equation}\label{FLRW3}
\frac{\ddot{a}}{a}=-H^{2}+\frac{1}{\alpha^{2}a^2}L^{2}-\frac{k}{a^2l^2}\, ,
\end{equation}
with
\begin{equation}\label{B}
B=H \, ,
\end{equation}
which gives $b(t)=(b_{_{\circ}}/a_{_{\circ}})a(t)\equiv\alpha a(t)$.
Throughout this paper the subscript `$_{\circ}$' is used to indicate the present value of the quantities.
In Eqs. (\ref{FLRW1}) and (\ref{FLRW2}) we have defined $\tilde{\rho}\equiv\rho_{_{M}}+\rho_{_{X}}$ and
$\tilde{p}\equiv p_{_{M}}+p_{_{X}}$ with energy density and pressure of dark energy as
\begin{equation}\label{Ro X}
\rho_{_{X}}\equiv T^{^{(X)}}_{_{tt}}=\frac{3}{8\pi G}\Big [\frac{1}{\alpha^2a^2}(L'+2L^2)-H^2 \Big ]
\end{equation}
and
\begin{equation}\label{P X}
p_{_{X}}\equiv -T^{^{(X)}}_{_{ii}}=\frac{1}{8\pi G}\Big[\dot{H}+3H^2-\frac{1}{\alpha^2a^2}(2L'+3L^2) \Big]\equiv w_{_{X}}\rho_{_{X}}\, .
\end{equation}

Since the detailed coupling form among the matter and dark energy is
unclear, one can expect their conservation equations may
not to be independent. Hence, the total energy conservation equation,
\begin{equation}\label{Conservation MIX}
\dot{\tilde{\rho}}+3H(\tilde{\rho}+\tilde{p})=0\, .
\end{equation}
can plausibly separate into two distinguished equations for $\rho_{_{X}}$ and
$\rho_{_{M}}$ as
\begin{equation}\label{Conservation X}
\dot{\rho}_{_{X}}+3H(\rho_{_{X}}+p_{_{X}})=f(t)
\end{equation}
and
\begin{equation}\label{Conservation M}
\dot{\rho}_{_{M}}+3H(\rho_{_{M}}+p_{_{M}})=-f(t)\, ,
\end{equation}
where $f(t)$ is assumed to be the interacting term between dark energy and matter.

By some manipulation, from Eqs. (\ref{Conservation X}), (\ref{FLRW1}) and (\ref{FLRW3}) (see Ref.~\cite{Amirfarshad2016}) one gets
\begin{equation}\label{f}
f(t)=H\rho_{_{M}}\, ,
\end{equation}
and
\begin{equation}\label{L}
L=\sqrt{C} \qquad {\rm and} \qquad L'=0\, ,
\end{equation}
which gives $l(y)=l_{\circ}e^{\pm \sqrt{C}(y-y_{\circ})}$ where $C$ is a positive real constant.
Eq. (\ref{f}) emphasizes that in this model, matter and dark energy do interact with each other.

By substituting Eq. (\ref{FLRW1}) in (\ref{FLRW3}) and considering (\ref{L}), one has
\begin{equation}\label{NFLRW3}
\frac{\ddot{a}}{a}=-\frac{4\pi G}{3}\rho_{_{M}}-\frac{k}{2l_{\circ}^2a^2}\, .
\end{equation}
Eq. (\ref{NFLRW3}) illustrates that, the geometrical term in acceleration equation
is capable to make the universe expansion accelerates, when, universe is open.
However, the measurements of anisotropies in the cosmic microwave
background suggest that the ordinary $4D$ universe is very close
to a spatially flat one~\cite{Bachcall}--\cite{Balbietal},
but it does not mean the universe is exactly spatially flat.
Recently, it has been shown that the observed large--scale cosmic microwave anomalies,
discovered by WMAP and confirmed by the Planck satellite, are most naturally explained in the context of a
marginally--open universe~\cite{Liddle}.
Also, there are several approaches, based on the string theory or quantum cosmology,
in favor of the open universe~\cite{Marina}--\cite{Coule2000}.
Therefore sine now, we concentrate on the spatially open universe, i.e. $k=-1$.

Combination of Eqs. (\ref{Conservation M}) and (\ref{f}) gets
\begin{equation}\label{Ro M}
\rho_{_{M}}=\rho_{_{M\circ}} \Big (\frac{a_{\circ}}{a}\Big)^4\, ,
\end{equation}
thus, Eq. (\ref{NFLRW3}) becomes
\begin{equation}\label{NFLRW3B}
\frac{\ddot{a}}{a}=-\frac{4\pi G}{3}\rho_{_{M\circ}} \Big (\frac{a_{\circ}}{a}\Big)^4 +\frac{1}{2l_{\circ}^2a^2}\, ,
\end{equation}
which shows that by the universe expansion the density of matter falls off more rapidly than the geometrical term.
Hence, the matter dominating (deceleration) epoch in early time cannot last for ever and dark energy, i.e. geometrical term, (acceleration)
finally comes to dominate. However, at the end, when scale factor $a$ becomes very very large, $\ddot{a}=0$ and universe will expand freely.
See Table~$1$ and Fig.~$1$.

\begin{figure}
\begin{center}
\epsfig{figure=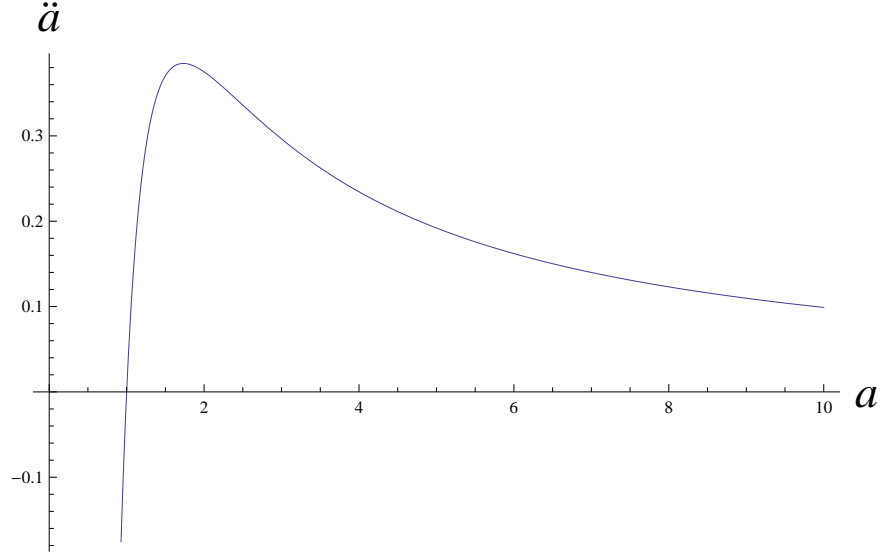} \caption{\footnotesize Changes of universe acceleration with
evolution of the scale factor of ordinary spatial dimensions. For better show, the scales of the axes has been resized by taking
$4\pi G\rho_{_{M\circ}}/3\equiv1\equiv2 l_{\circ}$, in Eq.~(\ref{NFLRW3B}) which gives $\ddot{a}=0$ when
$a=a_{\circ}^2 l_{\circ}\Big(8\pi G\rho_{_{M\circ}}/3\Big)^{1/2}\equiv1$.}
\end{center}
\end{figure}
\begin{table}
\begin{center}
\begin{tabular}{|c|c|c|c|}
  \hline
$\rm Universe \ Phase$ & $\ddot{a}$ & $a$  \\ \hline\hline
$ $ & $ $ & $ $ \\ 
$\rm Deceleration$ & $\ddot{a}\leq0$ & $0<a\leq a_{\circ}^2 l_{\circ}\Big(8\pi G\rho_{_{M\circ}}/3\Big)^{1/2}$  \\ 
$ $ & $ $ & $ $ \\ \hline
$ $ & $0<\ddot{a}< \ddot{a}_{_{Max}} $ & $a_{\circ}^2 l_{\circ}\Big(8\pi G\rho_{_{M\circ}}/3\Big)^{1/2}<a< a_{\circ}^2 l_{\circ}\Big(8\pi G\rho_{_{M\circ}}\Big)^{1/2}$  \\ 
$\rm Accceleration$ & $\ddot{a}=\ddot{a}_{_{Max}}=\frac{1}{3}a_{\circ}^{-2} l_{\circ}^{-3}\Big(8\pi G\rho_{_{M\circ}} \Big)^{-1/2}$ & $ a= a_{\circ}^2 l_{\circ}\Big(8\pi G\rho_{_{M\circ}}\Big)^{1/2}$ \\ 
$ $ & $\ddot{a}_{_{Max}}>\ddot{a}>0$ & $ a_{\circ}^2 l_{\circ}\Big(8\pi G\rho_{_{M\circ}}\Big)^{1/2}<a<\infty$ \\ \hline
\end{tabular}
\end{center}
\caption{\footnotesize Deceleration and acceleration epochs of the universe expansion and corresponding
scale factor of ordinary spatial dimensions.}
\end{table}

By comparing Eqs. (\ref{FLRW2})--(\ref{P X}) and (\ref{NFLRW3}) and some manipulations one can get the state parameter of dark energy,
$w_{_{X}}$ as
\begin{equation}\label{WX}
w_{_{X}}=-\frac{1}{3}\Bigg (1+\frac{1}{l_{\circ}^{2}(2Ca_{\circ}^{2}/b_{\circ}^{2}-\dot{a}^{2})}\Bigg)\, .
\end{equation}
While after the Big Bang in classical limit, the speed of universe expansion was too high therefore,
according to the relation (\ref{WX}), the limit of state parameter was, $w_{_{X}}=-1/3$.
Then, matter domination caused universe expansion to slow down until
$\ddot{a}=0$ when, $a=a_{\circ}^2 l_{\circ}\Big(8\pi G\rho_{_{M\circ}}/3\Big)^{1/2}$
and $w_{_{X}}=-Ca_{\circ}^{2}l_{\circ}^{2}\Big/3\Big (Ca_{\circ}^{2}l_{\circ}^{2}-b_{\circ}^{2}\Big)$.
After that, by the dark energy domination, $\ddot{a}>0$, which according to the Fig.~$1$ and Table~$1$
has a maximum, $\ddot{a}_{_{Max}}=1\Big/3a_{\circ}^{2} l_{\circ}^{3}\Big(8\pi G\rho_{_{M\circ}} \Big)^{1/2}$ at
$a= a_{\circ}^2 l_{\circ}\Big(8\pi G\rho_{_{M\circ}}\Big)^{1/2}$ with
$w_{_{X}}=-\Big(3Ca_{\circ}^{2}l_{\circ}^{2}+b_{\circ}^{2}\Big)\Big/3\Big(3Ca_{\circ}^{2}l_{\circ}^{2}-2b_{\circ}^{2}\Big)$.
Since then, the universe acceleration becomes small and small and finally get zero which causes universe expands freely
when, $w_{_{X}}=-\Big(2Ca_{\circ}^{2}l_{\circ}^{2}+b_{\circ}^{2}\Big)\Big/3\Big(2Ca_{\circ}^{2}l_{\circ}^{2}-b_{\circ}^{2}\Big)$.

At the end, we should remind that our approach in this manuscript is different from the brane world
scenarios~\cite{Randall:1999ee}--~\cite{Maartens:2010ar} and is based on the idea of induced matter
theory, which have been applied for a non--vacuum five--dimensional version of general relativity,
to introduce a geometrical interpretation for dark energy. In another word, in this model,
the source of dark energy in the $4D$ space--times can be viewed as a manifestation of
extra dimension which imposes a geometrical term containing the intrinsic curvature,
i.e. $k$, in the acceleration equation. Such a mechanism do not exist in the brane models.
\section{Conclusions}
\indent
It is a general belief that, what usually called dark energy makes the universe expansion to accelerate.
Though, an enormous amount of work has been performed to explain this acceleration,
but in many, the origin and the nature of dark energy is unclear yet.

Following the approach of the induced matter theory, we have investigated the cosmological
implications of a non--vacuum five--dimensional version of general relativity in order to explain both
decelerating and accelerating eras of the universe expansion. In this regard, on a $4D$ hypersurface,
we have classified the energy--momentum tensor into two parts.
One part represents all kind of the matter (the baryonic and dark) and the other one contains every
extra terms emerging from the fifth dimension, and has been considered as dark energy.
Afterwards, in a $5D$ space--time we have considered a generalized FLRW metric
and derived the FLRW type equations and also energy conservation equation on the $4D$ hypersurface.

Investigating the cosmological equations shows that matter (baryonic and dark) and dark energy interact with each
other and also the geometry of the universe contributes in acceleration equation.
The consistency of the model with observations imposes that the universe must be spatially open.
Generally, the model is capable to explain respectively the decelerated expansion in early time and then
accelerated expansion in late time of the universe evolution. However, it also predicts, this acceleration
has a maximum value and after that decreases slowly to zero at infinity which,
makes the universe expands freely at last.

\section*{Acknowledgements}
\indent

We would like to thank Prof. Tim M.P. Tait at University of California, Irvine, for reading this article and useful comments.


%
\end{document}